\documentclass[conference]{IEEEtran}
\IEEEoverridecommandlockouts
\usepackage{cite}
\usepackage{amsmath,amssymb,amsfonts}
\usepackage{algorithmic}
\usepackage{graphicx}
\usepackage{textcomp}
\usepackage{xcolor}
\usepackage[utf8]{inputenc}
\usepackage{fourier}
\usepackage{array}
\usepackage{makecell}
\usepackage{subfig}
\usepackage{hyperref}
\usepackage{cite}

\def\BibTeX{{\rm B\kern-.05em{\sc i\kern-.025em b}\kern-.08em
    T\kern-.1667em\lower.7ex\hbox{E}\kern-.125emX}}
\begin{document}

\title{Constrained-optimization Approach Delivers Superior Classical Performance for Graph Partitioning via Quantum-ready Method\\
}

\author{
\IEEEauthorblockN{Uchenna Chukwu}
\IEEEauthorblockA{\textit{Quantum Computing Inc.} \\
Leesburg, VA, USA \\
ORCID 0000-0002-1311-3827}
\and
\IEEEauthorblockN{Raouf Dridi}
\IEEEauthorblockA{\textit{Quantum Computing Inc.} \\
Leesburg, VA, USA \\
}
\and
\IEEEauthorblockN{Jesse Berwald}
\IEEEauthorblockA{\textit{Quantum Computing Inc.} \\
Leesburg, VA, USA \\
ORCID 0000-0003-4741-2427}
\and
\IEEEauthorblockN{Michael Booth}
\IEEEauthorblockA{\textit{Quantum Computing Inc.} \\
Leesburg, VA, USA \\
}
\and
\IEEEauthorblockN{John Dawson}
\IEEEauthorblockA{\textit{Quantum Computing Inc.} \\
Leesburg, VA, USA \\
ORCID 0000-0001-7665-5749}
\and

\IEEEauthorblockN{DeYung Le}
\IEEEauthorblockA{\textit{Quantum Computing Inc.} \\
Leesburg, VA, USA \\
}
\and
\IEEEauthorblockN{Mark Wainger}
\IEEEauthorblockA{\textit{Quantum Computing Inc.} \\
Leesburg, VA, USA \\
}
\and
\IEEEauthorblockN{Steven P. Reinhardt}
\IEEEauthorblockA{\textit{Quantum Computing Inc.} \\
Leesburg, VA, USA \\
ORCID 0000-0003-4355-6693}
\thanks{Corresponding author: sreinhardt@quantumcomputinginc.com}
}

\maketitle

\begin{abstract} 

Graph partitioning is one of an important set of well-known compute-intense (NP-hard)  graph problems that devolve to discrete constrained optimization. 
We sampled solutions to the problem via two different quantum-ready methods 
to understand the strengths and weaknesses of each method.
First we formulated and sampled the problem as a quadratic unconstrained binary optimization (QUBO) problem, via the best known QUBO formulation, using a best-in-class QUBO sampler running purely classically.  
Second, we formulated the problem at a higher level, as a set of constraints and an objective function, and sampled it with a recently developed constrained-optimization sampler 
(which internally generates and samples the problem via QUBOs also sampled classically).  
We find that both approaches often deliver better partitions than the purpose-built classical graph partitioners.  
Further, we find that the constrained-optimization approach is often able to deliver better partitions in less time than the bespoke-QUBO approach, without specific knowledge of the graph-partitioning problem.

Stepping back from graph partitioning itself, one key controversial question 
is whether bespoke algorithms for high-value problems 
or general tools for a class of problems are more likely 
to deliver the power of QCs to a broad market of real-world users.   
These results bear on that question, though they only use a few instances 
and require confirmation on other problems and other instances
as well as replacement of the low-level sampler 
by a QC instead of a classical software sampler.  
Still, this early evidence supports the proposition that 
general tools may contribute significant 
benefit to a range of problems, expanding the impact of QCs on industry and society.  
The fact that this benefit is independent of the low-level sampler employed, 
whether classical software or one of a variety of QC architectures, 
reinforces the need for further work 
on high-level optimization.  
The commercial availability in the cloud of such software today, delivering 
superior classical performance 
for some problems, enables quantum-forward organizations 
to migrate to quantum-ready methods now, 
accelerating progress toward quantum advantage and 
ensuring sampler software evolves
to address the problems such organizations value.

\end{abstract}


\section{Introduction}

The analysis of large graphs has become a valuable tool in an expanding set of domains where researchers seek to understand biological, social, cybersecurity, logistics, and other phenomena.  Because of the prohibitive computational time required, researchers have been effectively precluded from using compute-intense (NP-hard) graph kernels on all but the smallest graphs. Nevertheless, due to the potentially high-value insights provided by those kernels, researchers continue to seek ways to use such kernels, even when they provide only approximate, near-optimal solutions. 

The pioneering work of Mniszewski, Ushijima-Mwesigwa, Negre, et al. \cite{mniszewski2016graph, ushijima2017partitioning, negre2019communities}  mapped graph partitioning and community detection, two of those NP-hard kernels, to an early D-Wave annealing-based quantum computer (QC) via the quadratic unconstrained binary optimization (QUBO) formulation.  They found that in some cases a hybrid classical-quantum solver using the D-Wave 2X\texttrademark\hspace{1pt} QC found superior partitions to the most widely used graph partitioning software, Metis \cite{karypis1998metis} and KaHiP \cite{sanders2013kahip}.  They also recognized the disconnect between the very large scale of real-world graphs and the relatively tiny scale of graph kernels that can be solved directly on current QCs and hence have focused recent work \cite{ushijima2019multilevel} on \textit{multilevel} algorithms that depend on the direct solution of subproblems and the combination of those subsolutions into the solution of the whole problem.  In the multilevel context, the size of practical subproblems and the quality of results for those subproblems are essential factors.

We extended that work to larger graphs partitioned via calls to a NetworkX-like \cite{hagberg2008networkx} API with an underlying QUBO sampler running purely classically, to see whether the QUBO formulation, when coupled with a powerful sampler, can give present-day advantage.  We find that the QUBO formulation sampled by the QCI qbsolv\texttrademark\hspace{1pt} software running purely classically often delivers better partitions than the best classical partitioners, while also delivering good diversity of near-optimal results, as measured by graphs from the Walshaw graph-partition repository\cite{walshaw2002graph}.

Next, we started from a simpler constrained-optimization formulation 
of the graph-partitioning problem, consisting 
of the constraints and an objective function, 
an approach that enables an extra class of problem optimizations, 
such as choosing a Lagrange multiplier that separates 
infeasible and feasible solutions. 
We built an iterative sampler that finds a near-optimal Lagrange multiplier. 
When applied to some of the graph instances sampled earlier with a bare QCI qbsolv, 
we found a significant improvement over the bespoke-QUBO results. 
We show that the constrained-optimization formulation 
finds better answers than the simple QUBO formulation 
while maintaining 0\% size imbalance between the partitions. 

These capabilities are commercially available in the 2.0 release 
of QCI's Mukai software-execution platform, which is quantum-ready and cloud-based,
and includes the QCI NetworkX graph-analysis package, the QuOIR constrained-optimization sampler and the QCI qbsolv QUBO sampler.

During the current period when QCs do not yet deliver superior quantum performance,
superior classical performance is an important milestone but needs to be 
buttressed by a strong path to execution on real QCs.  
Qbsolv has previously demonstrated its ability to run both in purely classical
and hybrid classical/quantum modes for annealing-based QCs, and
the Quantum Approximate Optimization Algorithm (QAOA) \cite{farhi2014quantum} has been demonstrated
to solve QUBOs effectively on gate-model QCs.
Thus, the path to execution of constrained-optimization problems on diverse 
QC architectures appears to be limited by engineering rather than science.
Additionally, the software implementation of the Lagrange sampler resembles 
that of QAOA, in that each of them mixes satisfying the constraints with finding
optimal samples, which is needed for any such method.  
This similarity facilitates the future integration 
of gate-model QCs in QCI's Mukai platform. 
As this work raises the level of problem expression, it opens 
a new vein of possible classical optimizations 
before presenting a problem to a quantum computer.

On a different plane, the community of computer developers and users 
who expect to benefit from QCs is struggling with how to shift 
from current methods and implementations that target classical computers 
to not-yet-established methods and implementations that 
exploit the astonishing anticipated computational power of QCs. 
This shift is complicated by the highly uncertain delivery date 
of quantum advantage for real-world problems.

\begin{table*}[htp]
\begin{center}
\begin{tabular}{|c|r|r|r|r||r|r|r||r|r|r|}
\hline
 \multicolumn{5}{|c||}{} & \multicolumn{3}{|c||}{\thead{Bespoke-QUBO}} & \multicolumn{3}{c|}{\thead{Constrained-optimization}} \\
\hline
\thead{Graph} & \thead{|V|} & \thead{|E|} & \thead{Smallest \\ previously \\ known cut}  & \thead{Cut from \\ reference \\ \cite{ushijima2017partitioning}} & \thead{cut} & \thead{diversity \\ (by cut)} & \thead{sample \\ time (s)} & \thead{cut} & \thead{diversity \\ (by cut)} & \thead{sample \\ time (s)} \\
\hline
add20   & 2,395	& 7,462	& 596  & 647    & \textcolor{teal}{595}   & \makecell[r]{ \textcolor{teal}{595 x3} \\ 596 \textcolor{teal}{x51}} & 32 & \textcolor{teal}{595} & x1 & -- \\
\hline
3elt    & 4,720 & 13,722 &	90  & 90    & 90 &  90 \textcolor{teal}{x77}    & 140   & 90 & \textcolor{teal}{x32} & 22 \\
\hline
bcsstk33 & 8,738    & 291,583	& 10,171 & 10,171 & \textcolor{teal}{10,162} & \makecell[r]{\textcolor{teal}{10,162} x1 \\ \textcolor{teal}{10,164 x2} \\ \textcolor{teal}{10,166} x1}  & 184   & \textcolor{teal}{10,162} & x1 & 11 \\
\hline
vibrobox & 12,328	& 165,250	& 10,343 & na    & na & na & na & \textcolor{teal}{10,334} & \textcolor{teal}{x18} & 3,040 \\
\hline
4elt & 15,606	& 45,878 & 139  & na    & 139   & x1    & 7,379 & 139 & x1 & 1,514 \\
\hline
cti & 16,840	& 48,323 & 334  & na    & 334   & x1    & 7,408 & 334 & \textcolor{teal}{x4} & 210 \\
\hline
memplus & 17,758    & 54,196	 & 5,499 & na    & \textcolor{red}{6,190} & x1    & 7,413 & \textcolor{red}{5,537} & x1 & 10,809 \\
\hline 
bcsstk30 & 28,924	& 1,007,284	& 6,394 & na    & \textcolor{teal}{6,389} & \makecell[r]{ \textcolor{teal}{6,389} x1 \\  \textcolor{teal}{6,391} x1  \\ 6,394 \textcolor{teal}{x2}} & 7,846 & \textcolor{teal}{6,375} & x1 & 2,013 \\
\hline 
\end{tabular}
\caption{Graph bipartitioning results for bespoke-QUBO and constrained-optimization formulations sampled by Mukai samplers compared to best previously known results.  The results are black where equal 
to best results prior to this work,
\textcolor{teal}{green} where Mukai results (cut size or diversity) are better, 
and \textcolor{red}{red} where Mukai results are worse.}
\label{tbl:walshaw_graphs}
\end{center}
\end{table*}

The establishment of quantum-ready problem formulations and software (in this case, constrained optimization) to map those formulations efficiently to both classical computers and near-term QCs enables quantum-forward users to start migrating their key problems to QCs today.  Quantum-ready formulations that deliver present-day performance advantages, even if small by the immense expectations for QCs, would provide a practical migration path for quantum-forward organizations.

\section{Methods}

\subsection{Bespoke-QUBO Formulation}
We use the graph-bipartitioning formulation from Ushijima-Mwesigwa et al. \cite{ushijima2017partitioning}, notably Equation 23, with the clarifications (based on private communications with the authors) that, when $\alpha$ is equal to $\beta$, $Q_{ij}$ is set to $g_i - 1$ (the degree of the vertex minus 1) and that off-diagonal QUBO elements are doubled to account for the QUBO matrix representation being upper triangular (and hence asymmetric) while the graph is symmetric. We set $\alpha = \beta = 1.0$ for these experiments. Note that this formulation results in a dense QUBO with the number of elements growing as the square of the number of variables in the QUBO, and so very large QUBOs become unwieldy in data size as well as taking longer to solve. This formulation was implemented in version 1.1 of Quantum Computing Inc.’s QCI NetworkX (QNX) package available as part of QCI’s Mukai product.

\subsection{Constrained-optimization Formulation}


Graph partitioning can also be implemented as a constrained-optimization problem,
where the constraints are that each vertex can be assigned to exactly one partition
and the objective function is to have as few inter-partition edges as possible
(i.e., a minimum \textit{cut size}).
When formulated this way, an important detail is choosing 
the Lagrangian multiplier (typically named $\alpha$ though with a different
meaning from the $\alpha$ in the formulation in \cite{ushijima2017partitioning})
so that \textit{infeasible} solutions (i.e., those not respecting constraints) are 
separated in objective-function values from \textit{feasible} solutions.
This Lagrange sampling method is implemented in the QuOIR constrained-optimization component of the Mukai 2.0 release.

\section{Experimental Design}

\subsection{Targeted Problems}
Mniszewski et al. use graphs from the Walshaw graph partitioning repository \cite{walshaw2002graph}, namely the add20, data, 3elt, and bcsstk33 graphs.  We used add20, 3elt, bcsstk33, vibrobox, 4elt, cti, memplus, and bcsstk30, targeting bigger problems. Mniszewski et al. used graphs up to 8,738 vertices and 291,583 edges; this work extends that considerably, with the largest graph (bcsstk30) consisting of 28,924 vertices and 1,007,284 edges. See Table \ref{tbl:walshaw_graphs}  for graph details. 

\subsection{Execution Context}
We solve the QNX-generated QUBOs with QCI qbsolv, another component of the Mukai software stack.  Derived from the open-source qbsolv \cite{booth2017qbsolv}, QCI qbsolv has been reimplemented to deliver exceptional performance (quality, speed, and diversity of results) from highly parallel classical processors \cite{booth2020qbsolv}, exploiting advanced tabu search techniques.  
These results were obtained running Mukai on an AWS instance visible via a Python API exercising a REST API.  
The bespoke-QUBO results were obtained on Mukai version 1.1 using a c5n.18xlarge AWS Linux instance, including 72 Intel Xeon cores running at 2.0 GHz and 192 GiB of RAM.
The constrained-optimization results were obtained on Mukai version 2.0 running 
on a c5.24xlarge AWS Linux instance, including 96 Intel Xeon cores running 
at 3.0 GHz (up to 3.4GHz via Turbo Boost) and 192 GiB of RAM.

\subsection{Key Metrics}
For graph partitioning, we define the quality of the results as the number of edges in the cut between the two partitions, where smaller is better.  
Diversity of results means the degree to which results 
with similar or equal cut sizes make markedly different assignments 
of vertices to partitions, with greater diversity being better \cite{lewis2017preprocessing}. 
Depending on the use case, the quality of the result, the speed 
of obtaining the result, and the diversity of results 
are important metrics for the graph-partitioning kernel.  
We provide all three in Table \ref{tbl:walshaw_graphs}.  
We targeted 0\% imbalance, consistent with \cite{ushijima2017partitioning} 
and the first results table of the Walshaw repository \cite{walshaw2002graph}.

The Walshaw repository lists results from 43 solvers,
of which 14 got the best result (smallest cut size) on at least 1
of the 34 graphs, for the 0\%-imbalance case.
Mniszewski et al. compared their results to Metis and KaHIP.
Table \ref{tbl:walshaw_graphs} compares Mukai results with the best
of either the Walshaw or Mniszewski results.

The Mukai results shown are the best of 5 executions.

\section{Results}
The results in Table \ref{tbl:walshaw_graphs} are black where equal 
to best results prior to this work,
\textcolor{teal}{green} where Mukai results (cut size) are better, 
and \textcolor{red}{red} where Mukai results are worse.

\subsection{Bespoke-QUBO Formulation}
As shown in Table \ref{tbl:walshaw_graphs}, for these problems the bespoke-QUBO formulation delivers cut sizes that are often better than the best previously known cuts and often equal, while in one case noticeably poorer. 
The value of a better cut, even if only moderately better, is context dependent. 
For multilevel graph partitioning, even a small improvement reduces 
the exponential growth of compute time at the next level and so can be high value.  
The bespoke-QUBO formulation delivers high diversity of excellent solutions, finding 3 solutions better than the previously best known for add20, and finding 4 distinct better solutions at 3 distinct smaller cut sizes for bcsstk33 and one solution at each of 2 distinct smaller cut sizes for bcsstk30. 
We denote this in the table by (e.g.) “595 x3” for add20, meaning that cut size 595 was found by 3 distinct solutions. We note that the smallest cut for bcsstk33 came from a solution whose energy (-19,032,362) was slightly less optimal than the energy (-19,032,364) of a solution with a larger (less optimal) cut; this bears more investigation. Diversity is often valuable, as multiple solutions can provide a view of the robustness of the solution and support human-in-the-loop or automated post-processing by criteria not captured by bipartitioning.  We also note that prior best Walshaw results \cite{walshaw2002graph} from the 8 graphs we use come from 4 different solvers (PROBE (add20), JE (3elt, vibrobox, 4elt, cti, and bcsstk30), GCSVD (bcsstk33), and *+ILP (memplus), so this set is stressing the diversity of solvers.

\subsection{Constrained-optimization Formulation}

Relevant results are represented in the last three columns of Table \ref{tbl:walshaw_graphs}. The general observation is that the constrained-optimization formulation continues to deliver high diversity and excellent quality, now, with much shorter times.  A noticeable example is cti where the sample time is substantially reduced from around 2 hours to only 201 seconds (with four times more diversity). This is also true for the other graphs with the exception of memplus which continues to be a challenging instance for both formulations.  We also note three significant improvements: memplus from 6,190 to 5,537 (though still above the smallest known cut),  vibrobox where  the cut-size is down to 10,334, with excellent diversity, and finally, bcsstk30, with cut-size equal to 6,375, 
further improvement beyond the bespoke-QUBO improved number, 
with time reduced to almost a quarter of the previous time.  
Surprisingly,  even with a fine-tuned choice of the Lagrange multiplier, the best cut-size 
for bcsstk33 also came from a suboptimal sample. 
We continue investigating this type of anomaly. 

With graph partitioning being an NP-hard problem, we expect the time to solution to grow considerably with problem size, for both formulations.
The drastic bespoke-QUBO growth from bcsstk33 to 4elt and the plateau from 4elt to bcsstk30 is unexplained and we are investigating further. Note that, for many use cases, the time spent calculating the best answer will be more important than the cost, which was O(\$5/hour) for cloud instances. The time to solution for problems up to bcsstk33 appears usable in a multilevel graph partitioning context where we would expect numerous repeated calls to partition subgraphs.  (The Walshaw repository does not record the execution time or environment for submitted results, so comparisons in those dimensions are not possible.)

\section{Discussion}

\subsection{Graph Partitioning Results}
To our knowledge, these are the first results from a concerted effort to deliver superior graph-bipartitioning performance with a quantum-ready formulation executing purely classically.  We find that, for the small set of graph instances studied, this implementation often delivers better cut sizes, in addition to strong diversity of solutions.  

Reference \cite{mniszewski2016graph} solved the same problem with the same formulation, targeting small graphs from the same repository, using a hybrid classical-quantum solver (open-source qbsolv) and reported sometimes finding better results than the most popular purpose-built graph partitioners. Contributions from classical versus quantum execution were not isolated.  This work extends \cite{mniszewski2016graph} by a) executing purely classically, b) extending the range of graphs to bigger instances, c) obtaining results better than the best known results (i.e., for partitioners beyond METIS and KaHIP), and d) measuring diversity of solution as well as quality of solution. 

The Walshaw problems, with ranges of size and density, present an excellent challenge for constrained-optimization and QUBO solvers.  Achieving these results markedly improved the performance of the QNX and QuOIR packages and the underlying Mukai components on which they depend.

We are not aware of graph-bipartitioning results from the most popular purpose-built classical partitioners (e.g., METIS and KaHIP) that normalize for the volume of compute resources devoted to their solution.  Having such results would isolate how much of the benefit shown here is due to a better formulation and how much is due to the use of more compute resources. 

\subsection{Alternative Implementations of Constrained-optimization Formulation}
Quantum-ready formulations and implementations are in their infancy, relative to the maturity of purpose-built classical implementations, and we expect rapid improvement in quantum-ready implementations.  
We are investigating various extensions. In particular, one possibility is to use the rich variety of samples returned by QCI qbsolv to perform an update 
instead of the single ground sample as currently implemented. 
These samples collectively constitute the lowest energy levels of the spectrum, 
and thus provide more information about the energy landscape, which ultimately  improves convergence speed and quality of solutions. 
Furthermore, it is common to view a QUBO  as a weighted graph, with node biases on the diagonal and edge weights specified by off-diagonal terms. Low-dimensional embeddings of a graph, such as the methods developed in {\em node2vec}~\cite{grover2016node2vec}, have proven  powerful for graph classification. This combination of constrained optimization in the form of a QUBO with machine learning and graph analysis is currently being investigated as a way to shorten the search time for optimal Lagrange multipliers. Another complementary direction we are pursuing is the use of Groebner bases, an approach in which the objective function is detached from the problem and only queried as an oracle, while all of the work is done at the constraints level \cite{dridi2019minimizing}. 

Since QCI NetworkX exposes a high-level graph API, app developers gain high ease of use while benefiting automatically from improvements to the underlying implementation.

An obvious benefit of decomposing tools like QCI qbsolv is that they enable solution of real-world problems today, so app developers can learn of most pitfalls to solving problems this way, and shed light on execution steps that are insufficiently fast today, enabling system developers to address them to deliver quantum advantage  earlier.

\subsection{Impact on Use of Near-term Quantum Computers}

Of course, the purely classical approach described here currently gains no benefit from quantum execution, as indeed few real-world problems do.  However, a quantum-ready classical sampler that often delivers superior performance is an excellent starting point to exploit quantum advantage when QCs deliver it.  The open-source qbsolv established that a partitioning QUBO solver can effectively solve large QUBOs, targeting a small-sized QPU.  The QCI qbsolv implementation improves quality of results and time to solution, and adds diversity by the return of multiple solutions. Reference \cite{ushijima2019multilevel} shows evidence (Figure 5) that solving bigger subproblems on bigger QPUs will give better subproblem contributions. The next-generation D-Wave system based on the Pegasus topology \cite{boothby2020nextgen,pelofske2019minvcov} is expected to grow the subproblem size to 180 variables compared with the current 64 (D-Wave 2000Q\texttrademark).  Performance of that system should be a major step toward quantum advantage. 
Reference \cite{hadfield2019qaoa} shows that QUBOs (subproblems, in the case of a decomposing QUBO sampler) can also be solved effectively on gate-model QCs
via QAOA, whose algorithm is similar to the Lagrange sampler 
described above. 
Thus, we expect that a combination of QuOIR algorithmic improvements, quantum-ready QUBO sampler improvements, extended with a strong decomposing ability (whether QCI qbsolv or a better alternative), and potential quantum advantage (from the next-gen D-Wave system or a future-model gate-model QC) will deliver the best possible performance available at that time. 

\section{Conclusions}
We report that a commercially available, quantum-ready execution platform, 
QCI's Mukai, supports two formulations of graph partitioning that often find  better solutions than the best previously known results.  
Further, the best results are obtained not from creating a higher quality QUBO
formulation but rather from expressing the problem in the language of constrained optimization -- constraints and an objective function -- and processing those
with advanced methods that will also apply to many other problem types.
Given the early state of quantum-ready formulations and solvers, we expect rapid improvements for such approaches.  
These early results show the potential for quantum-forward organizations to shift key computational components to a quantum-ready approach now, reaping superior classical performance now while preparing for an eventual migration to quantum computers delivering quantum advantage.

\section*{Acknowledgment}

The authors thank Susan Mniszewski and Hayato Ushijima-Mwesigwa for clarifications of the details of the QUBO formulation in [2].






\bibliographystyle{unsrt}
\bibliography{QuOIR_graph_partitioning_autotuning_performance}

\end{document}